\documentclass[conference]{IEEEtran}
\IEEEoverridecommandlockouts

\usepackage{cite}
\usepackage{amsthm,amsmath,amssymb,amsfonts}
\usepackage{graphicx}
\usepackage{url}
\usepackage{textcomp}
\usepackage{verbatim}
\usepackage{xcolor}
\usepackage{float}
\usepackage{tabularx}
\usepackage{listings}
\usepackage{hyperref}
\usepackage{array}
\usepackage{multirow}
\definecolor{verylightgray}{rgb}{.97,.97,.97}
\newcolumntype{P}[1]{>{\centering\arraybackslash}p{#1}}

\lstdefinelanguage{Solidity}{
	keywords=[1]{anonymous, assembly, assert, balance, break, call, callcode, case, catch, class, constant, continue, constructor, contract, debugger, default, delegatecall, delete, do, else, emit, event, experimental, export, external, false, finally, for, function, gas, if, implements, import, in, indexed, instanceof, interface, internal, is, length, library, log0, log1, log2, log3, log4, memory, modifier, new, payable, pragma, private, protected, public, pure, push, require, return, returns, revert, selfdestruct, send, solidity, storage, struct, suicide, super, switch, then, this, throw, transfer, true, try, typeof, using, value, view, while, with, addmod, ecrecover, keccak256, mulmod, ripemd160, sha256, sha3}, 
	keywordstyle=[1]\color{blue}\bfseries,
	keywords=[2]{address, bool, byte, bytes, bytes1, bytes2, bytes3, bytes4, bytes5, bytes6, bytes7, bytes8, bytes9, bytes10, bytes11, bytes12, bytes13, bytes14, bytes15, bytes16, bytes17, bytes18, bytes19, bytes20, bytes21, bytes22, bytes23, bytes24, bytes25, bytes26, bytes27, bytes28, bytes29, bytes30, bytes31, bytes32, enum, int, int8, int16, int24, int32, int40, int48, int56, int64, int72, int80, int88, int96, int104, int112, int120, int128, int136, int144, int152, int160, int168, int176, int184, int192, int200, int208, int216, int224, int232, int240, int248, int256, mapping, string, uint, uint8, uint16, uint24, uint32, uint40, uint48, uint56, uint64, uint72, uint80, uint88, uint96, uint104, uint112, uint120, uint128, uint136, uint144, uint152, uint160, uint168, uint176, uint184, uint192, uint200, uint208, uint216, uint224, uint232, uint240, uint248, uint256, var, void, ether, finney, szabo, wei, days, hours, minutes, seconds, weeks, years},	
	keywordstyle=[2]\color{teal}\bfseries,
	keywords=[3]{block, blockhash, coinbase, difficulty, gaslimit, number, timestamp, msg, data, gas, sender, sig, value, now, tx, gasprice, origin},	
	keywordstyle=[3]\color{violet}\bfseries,
	identifierstyle=\color{black},
	sensitive=false,
	comment=[l]{//},
	morecomment=[s]{/*}{*/},
	commentstyle=\color{gray}\ttfamily,
	stringstyle=\color{red}\ttfamily,
	morestring=[b]',
	morestring=[b]"
}

\usepackage[linesnumbered,ruled,vlined]{algorithm2e}

\def\BibTeX{{\rm B\kern-.05em{\sc i\kern-.025em b}\kern-.08em
    T\kern-.1667em\lower.7ex\hbox{E}\kern-.125emX}}

\usepackage{tabularx}

\newcounter{protocol}

\begin{document}

\title{Off-chain Execution and Verification of Computationally Intensive Smart Contracts
\thanks{Research supported by NSF awards \#1800088, \#2028797, \#1914635, Intel Labs, and the Federal Aviation Administration (FAA). Any opinions, findings, and conclusions or recommendations expressed in this material are those of the author(s) and do not necessarily reflect the views of the NSF, FAA, and Intel Inc.}
}              
                 
\author{\IEEEauthorblockN{Emrah Sariboz,
Kartick Kolachala, Gaurav Panwar, 
Roopa Vishwanathan, and Satyajayant Misra}                       
                                     
\IEEEauthorblockA{\textit{Department of Computer Science} \\
\textit{New Mexico State University}\\
Las Cruces, NM, USA \\
\{emrah, kart1712, gpanwar, roopav, misra\}@nmsu.edu}}
\IEEEoverridecommandlockouts
\IEEEpubid{\makebox[\columnwidth]{978-0-7381-1420-0/21/\$31.00~\copyright2021 IEEE \hfill} \hspace{\columnsep}\makebox[\columnwidth]{ }}
\maketitle

\IEEEpubidadjcol

\begin{abstract}
We propose a novel framework for off-chain execution and verification of computationally-intensive smart contracts. Our framework is the first solution that avoids duplication of computing effort across multiple contractors, does not require trusted execution environments, supports computations that do not have deterministic results, and supports general-purpose computations written in a high-level language. Our experiments reveal that some intensive applications may require as much as 141 million gas, approximately 71x more than the current block gas limit for computation in Ethereum today, and can be avoided by utilizing the proposed framework. 
\end{abstract}


\section{Introduction}
A smart contract is a computer program that resides on the Ethereum blockchain and gets executed automatically when predetermined conditions are met. 
Depending on the complexity, every transaction that modifies a smart contract's state consumes a certain amount of gas (the unit of cost in the Ethereum blockchain). 
As a result of this, it becomes infeasible to use smart contracts for computationally intensive applications such as image recognition and zero-knowledge proofs. In this paper, we refer to such contracts as \emph{computationally intensive smart contracts} (CICs).

Recent studies have explored alternative solutions to eliminate the cost and make CIC execution scalable. Proposed solutions to this end either replicate the CIC's execution across a small subset of nodes or require a Trusted Execution Environment (TEE), which engenders greater trust. 
An alternative to the aforementioned methods is to outsource the CIC computation to a third party that does the computation and generates a proof of correctness for the same, that can be verified in polynomial time. 
Using this approach, the client can verify the returned computation's correctness in a much more efficient manner than re-executing it.
Our work falls into verifiable computation category where we propose a solution that is scalable, avoids duplicating computations, and does not require tamper-resistant hardware or trusted execution environments.

\vspace{-0.15cm}
\section{Related Work}
\label{litrev}
%
%


%
%
\noindent \textbf{Trusted Hardware:} TEE has been adopted to alleviate scalability and confidentiality obstructions of smart contracts in {\cite{das2019fastkitten,cheng2018ekiden,wustace}}. However, recent studies have identified several attack on SGX --- we avoid the impact as we do not need SGX\cite{hahnel2017high,moghimi2017cachezoom,
van2017telling,lee2017inferring,xu2015controlled,nilsson2020survey,
gotzfried2017cache}. \textbf{ Replicated Computation:} Outsourcing CIC execution to a set of delegators has been proposed by \cite{das2018yoda}; however, this model suffers from the large overhead of replicated computation and lacks support for randomized computations, which we address in our work. \textbf{Verifiable Computation:} Interactive proofs (IPs)~ \cite{feige1996interactive} and probabilistically checkable proofs (PCPs)~\cite{arora1998probabilistic} laid the foundations of provable verifiable computation which has been studied in \cite{fc01,dg05,sion05,gm01,cmt12,muggles,parno10,vadhan10}.  Despite promising {\color{black} asymptotics}, these proof systems are highly impractical and may take inordinately long to verify instances with small input sizes~\cite{parno2013pinocchio}. 
Another line of work applies the above theoretical foundations to practice on cloud computing settings studied in \cite{pepper,ginger, allspice}; however, they are far from being scalable for general-purpose computation. Adoption of zk-SNARKs\cite{zksnark} to verify smart contracts was proposed by \cite{zokrates}; however, their solution requires that the application code be written in a domain-specific language that they designed. This differs from our work as our work supports computations that are written in a high-level language.

\begin{figure*}
	\begin{center}
		\includegraphics[width=0.71\textwidth]{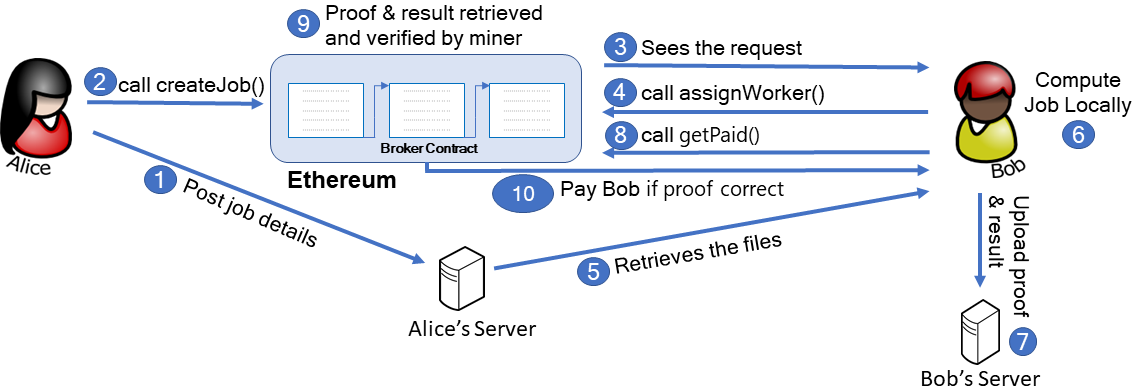}
		\caption{Schematic diagram of interactions between the entities and the corresponding function calls in the framework.}
		\label{fig:fig1}
	\end{center}
\end{figure*}

\section{Construction}

\par 
The components of our framework are as follows: a client (Alice) who wishes to outsource a computationally intensive job, a worker (Bob) who does the computation for the client in exchange for some monetary reward, a miner (Charlie) to validate the transactions, and \emph{Broker contract}, a smart contract which acts as an intermediary between the client and the worker. 

{\bf Client's Operations:} 
 Alice writes the details of the smart contract to be executed to her publicly accessible  server Step 1 in Figure~\ref{fig:fig1}. The details contain the inputs needed for execution, the fee given to a worker, the collateral the worker needs to deposit to register for this job, and the maximum time she is willing to allot for the computation result to be delivered to her. She posts this job creation request to the blockchain~ by interacting with the \emph{Broker Contract} in Step 2. This request contains the URL of her server, which has all the aforementioned details.

{\bf Worker's Operations:}  If Bob is interested in executing the computation, he goes to the specified server URL to check whether he has enough resources to complete the computation within the requested time interval in Step 3 and   registers for the job by depositing the required collateral to \emph{Broker Contract} in Step 4. He retrieves the inputs needed for the computation from Alice's server in Step 5. 
He performs the computation locally in Step 6, generates proof of correctness, and uploads them to his server in Step 7. He then submits the URL to \emph{Broker Contract} for the verification by calling \emph{getPaid()} function in Step 8 to get compensated for his work which internally starts the proof-verification mechanism. 

{\bf Miner's Operations:}
Charlie picks up the transaction posted by Bob, executes the \emph{Broker Contract}, and retrieves the result and proof from Bob’s server in Step 9. The \emph{Broker Contract} checks whether the proof was posted within a specified time limit, verifies the proof and result, and posts them to Alice’s server. \emph{Broker Contract} outputs a transaction paying Bob his fee and refunding his collateral if the verifications are successful in Step 10. However, Alice gets refunded her fee and also gets Bob’s collateral if the verifications fail.

\begin{table}[H]
	\centering
	\caption{The mean and standard deviation for computationally intensive applications}
	\label{tbl:exp12}
	\begin{tabular}{|P{1.65cm}|P{0.90cm}|P{1.35cm}|P{1.48cm}|P{1.35cm}|}
		\hline
		Computation                                                                        & Input          & KeyGen (s)       & ProofGen (s)      & Verify (ms) \\ \hline
		\multirow{2}{*}{
		\begin{tabular}[c]{@{}l@{}}Matrix Mult.\end{tabular}
		}  
		& 70$\times$70   & 40.25$\pm$1.48  & 117.91$\pm$4.24  & 2$\pm$2    \\ \cline{2-5} 
		& 110$\times$110 & 158.43$\pm$8.89 & 487.46$\pm$93.50 & 9$\pm$10   \\ \hline
		\multirow{2}{*}{\begin{tabular}[c]{@{}l@{}}Image Match.\end{tabular}}
		& 45$\times$45   & 32.23$\pm$0.76  & 75.12$\pm$1.89   & 70$\pm$61  \\ \cline{2-5} 
		& 85$\times$85   & 115.89$\pm$3.32 & 317.88$\pm$8.68  & 9$\pm$2    \\ \cline{2-5} 
		\hline
		
		\multirow{2}{*}{MultiVar Poly
		} 
		& 500040       & 36.23$\pm$2.58  & 140.81$\pm$7.12  & 8$\pm$2 \\ \cline{2-5} 
		& 644170       & 65.70$\pm$3.60  & 1220.82$\pm$8.59 &8$\pm$2 \\ \hline
		
		\multirow{2}{*}{Floyd-Warshall} & 16$\times$16 & 45.57$\pm$3.00  & 112.06$\pm$6.86  & 1$\pm$2  \\ \cline{2-5} 
		& 25$\times$25 & 166.40$\pm$4.36 & 514.99$\pm$13.35 & 3$\pm$7  \\ \hline
		
	\end{tabular}
\end{table}

\section{Results And Evaluation}

\par 
The proposed framework's performance has been evaluated on four computationally intensive applications as in~\cite{parno2013pinocchio}.
 
\textbf{Matrix Multiplication} takes two $n \times n$ matrices as an input, $M_1$ and $M_2$, and computes $M_1 \cdot M_2$.
\textbf{Image Matching} takes a $k_w \times k_h$ ($k_w = k_h = 3$) sized image kernel and computes the point in an image where the minimum difference happens between the image and the kernel. 
\textbf{Multi-Variate polynomial evaluation} takes a polynomial of degree $m$, containing $(m+1)^k$ coefficients, and evaluates it over $k$ ($k = 5$) variables taken as inputs. 
\textbf{Floyd-Warshall} algorithm takes an $n\times n$ matrix representing the adjacency matrix of an $n$-vertex graph. 
It computes the shortest paths among all the vertices. 

According to our calculations, the gas required to implement these applications in smart contracts is infeasible given the current block gas limit of $\approx$12 million \cite{ethstats}, e.g., 142 million gas units for image matching.

\textbf{Evaluation}:
The above applications were written in C and first transformed into an arithmetic circuit. Next, the Pinocchio compiler is used to generate Quadratic Arithmetic Program (QAP), evaluation and verification keys~\cite{parno2013pinocchio}. In our experiments, the key generation phase is completed by the client and given to the worker along with the QAP. On receiving these parameters, the worker executes the code and posts the proof to the server he controls.

%
%

Our experimental results, detailed in Table~\ref{tbl:exp12}, show that our framework provides quick proof verification, for different sizes of input parameters for all applications.
The framework also maintains a constant proof size of 288 bytes in all cases.
%
As expected, we have noticed an increase in the proof generation time with an increase in the application parameters' size. This growth was linear for all except image matching, which was super linear due to an increase in the number of multiplication gates and equality comparisons in the equivalent arithmetic table. 

\section{Conclusion}
\par We proposed a novel framework for execution and the verification of the CICs by offloading them to a computationally powerful entity using an incentive mechanism. Unlike other proposed solutions, our work prevents replicated computation, eliminates the need for TEEs, and supports computations with random results. 

\vspace{12pt}

\bibliographystyle{IEEEtran}
\bibliography{references}

\end{document}